\documentstyle[prd,eqsecnum,aps]{revtex}

\def\be{\begin{equation}}
\def\ee{\end{equation}}
\def\ba{\begin{eqnarray}}
\def\ea{\end{eqnarray}}

\def\negenspace{\kern-1.1em}

\def\semidirect{\;{\rlap{$\subset$}\times}\;}
\begin{document}
\draft


\title{Gravitational Goldstone fields from affine gauge 
theory}

\author{Romualdo Tresguerres\thanks{Email address: ceef310@imaff.cfmac.csic.es} \\
IMAFF, Consejo Superior de Investigaciones Cient\'{\i}ficas,\\
Serrano 113 bis, Madrid 28006, SPAIN\\
and\\
Eckehard W. Mielke$^\diamond$\thanks{Email address: ekke@xanum.uam.mx} \\
$^{\diamond}$ Departamento de F\'{\i}sica,
Universidad Aut\'onoma Metropolitana--Iztapalapa,\\
Apartado Postal 55-534, C.P. 09340, M\'exico, D.F., MEXICO}
\maketitle

\today

\begin{abstract}

In order to facilitate the application of standard renormalization 
techniques, gravitation should be decribed, if possible, in pure 
connection formalism, as a Yang--Mills theory of a 
certain spacetime group, say the 
Poincar\'e or the affine group. This embodies the translational 
as well as the linear connection. However, the coframe 
is not the standard Yang--Mills type gauge field of the translations, 
since it lacks the inhomogeneous gradient term in the gauge 
transformations. By explicitly restoring the ``hidden" piece responsible for this behavior within the framework of {\em nonlinear realizations}, the usual geometrical interpretation of the dynamical theory becomes 
possible, and in addition one can avoid the metric or coframe 
degeneracy which would otherwise interfere with the integrations 
within the path integral. We claim that nonlinear realizations 
provide a general mathematical scheme clarifying the foundations of gauge 
theories of spacetime symmetries. When applied to construct the 
Yang-Mills theory of the affine group, tetrads become identified 
with nonlinear translational connections; the anholonomic metric 
does not constitute any more an independent gravitational potential, 
since its degrees of freedom reveal to correspond to eliminable {\em Goldstone} bosons. This may be an important advantage for quantization.
\end{abstract}

\section{\bf Introduction}
On a macroscopic scale, gravity is empirically rather
well described by Einstein's general relativity theory (GR). However, quantum-field theoretically, Einstein's theory is perturbatively {\em 
nonrenormalizable}\cite{Is91,LN90} and plagued by {\em anomalies} 
if coupled to fundamental matter as the Dirac field of an 
electron, cf. \cite{MK98,KM00}. As a matter of fact, in the standard GR theory, gravity is conceived as an interaction 
of a very different nature as the remaining forces, being supposed 
to be mediated by the spacetime metric rather than by Yang--Mills 
connections. Thus, it is reasonable to hope that some of the 
problems related to the quantization of this force might disappear 
if one were able to describe gravitation as an ordinary 
gauge theory. The search for such a formulation yielded in particular 
the different gauge theories of gravity proposed by Hehl and his 
Cologne group \cite{HHKN76,He80,hehl95}. In all of them, the 
local treatment of translations reveals itself as a corner-stone 
of the Yang--Mills type interpretation of gravitation. As  Feynman \cite{Fe62} has put it, ``...gravity is that field which corresponds to a 
gauge invariance with respect to displacement transformations", 
cf. \cite{Mi92}. Starting from the Metric-Affine Gravity (MAG) theory \cite{hehl95}, our aim is to re-examine it from the point of view of nonlinear realizations (NLRs). We think that such approach enlightens several internal features of the theory. The interpretation of a few specific points becomes modified, but the main features of the formalism elaborated in the work by Hehl et al. \cite{hehl95}, in particular the 
field equations, remain unchanged under our alternative derivation.

\subsection{Tetrads as nonlinear connections}
In a first order formalism, one introduces a {\em local frame} field
(or vielbein), $e_{\alpha} = e^{i}{}_{\alpha}\,\partial_{i}\, $ 
together with the {\em coframe} field or one--form basis 
$\vartheta^{\beta} = e_{j}{}^{\beta}\,dx^{j}\>,$ which is dual to 
the frame $e_\alpha$ with respect to the {\it interior product}: 
$e_{\alpha}\rfloor\vartheta^{\beta} = \delta_{\alpha}^{\beta}\>$. 
Quite often, $\vartheta^{\alpha}$ is advocated as the translational 
gauge potential, although it does {\em not} transform inhomogeneously 
under local frame rotations, as is characteristic for a connection. 
The rigourous explanation of this apparent paradox requires to 
invoke a {\em nonlinear realization} (NLR) of the local spacetime group. 
In this paper, we will demonstrate that $\vartheta^{\alpha}$ actually is 
the dimensionless nonlinear translational gauge potential in 
Metric--Affine Gravity, that is, in the Yang--Mills 
approach to the affine group. (This quite general group includes 
the Poincar\'e group of elementary particle physics as subgroup.) 

As a consequence of the nonlinear treatment, the metric tensor 
will reveal itself as dynamically irrelevant, no more playing the 
role of a gravitational potential, since the degrees of freedom of 
the MAG--metric become rearrangeable into redefined connections. Although this fact is 
derived here independently, the idea of regarding the metric as 
the gravitational analogon of the Higgs or Goldstone field is not new. 
It has first been proposed by Isham, Salam and Strathdee \cite{ISS71} 
and lateron was occasionally discussed  by Nambu \cite {N78}, Ne'eman 
and Regge \cite{NR78}, and Trautman \cite{Tr79}, e.g.  Various 
existing models of macroscopic gravity can therefore be reinterpreted as nonlinear gauge theories, cf. \cite{Pi80,NFD80,Sm87,Mi87}. 
Lateron, Flato and R\c aczka \cite{FR88} as well as  van 
der Bij \cite{V96} speculated about a {\em gravitational origin} 
of the Higgs field and its decoupling.

\subsection{Spacetime metric as a Goldstone boson}

The recent paper by Gronwald et al. \cite{GMMH98} adopts the 
 ``quartet" of scalar fields introduced by Guendelman and 
Kaganovich\cite{Guendel1996a,Guendel1997a,Guendel1997b,Guendel1999a,Guendel1999b,Guendel1999d,Guendel2000}, 
originally used in 
the study of the cosmological constant problem. Gronwald et al. apply 
these fields in a different context, in order to remedy, as 
far as possible, the insatisfactory fact of the occurrence of 
the (anholonomic) metric tensor in MAG as a field different from 
the Yang-Mills ones. As declared in the introduction by the authors 
themselves, they look for an alternative way to define a 
volume element without reference to the metric. 

In fact the introduction, besides the linear connection and the
coframe, of the metric tensor as an independent gravitational
potential \cite{hehl95} seems to be contrary to the spirit of a pure 
gauge approach, where the role of gauge potentials is played 
exclusively by connections. In standard Yang--Mills theories, no 
other quantity is required to carry interactions. Contrarily, in MAG, 
the metric tensor appears as a quantity with ten additional 
degrees of freedom, foreign to the otherwise standard 
Yang--Mills treatment. 

In our opinion, nonlinear realizations \cite{ISS71} provide a
different interpretation of the actually formulated theory in its present form\cite{hehl95}. In fact, most of the features of MAG --- such as the vector character of the coframe despite its nature of (translational) connection, as already mentioned; the freedom to fix the metric tensor to be Minkowskian without loss of generality, the tensoriallity of the nonmetricity, etc. --- are consequences of the 
nonlinearity. For instance, as we will see, the Goldstone nature of the MAG metric highlights from a particular nonlinear realization of the affine group. According to this interpretation, the metric results to be a set of ten fields, rearrangeable in the connections. When rearranged 
in this way, the metric reduces to a constant one, without 
any dynamical degrees of freedom. Thus, the anholonomic MAG metric 
tensor is not a genuine gravitational potential. Accordingly, the 
nonmetricity is not to be interpreted as the corresponding field 
strength but simply as the connection component associated to the 
symmetric generators of the general linear group. Under 
gauge transformations, nonmetricity behaves as a tensor due 
to the (not immediatly recognizable) underlying nonlinear 
realization of the affine group.
\bigskip

\section{Gauging spacetime groups}
As far as internal symmetries are concerned, the definition of
gauge transformations as fibre-preserving bundle 
automorphisms \cite{AHS78} constitutes a satisfactory characterization of them. 
Accordingly, given a principal fibre bundle $P\left( M\,,H\,
\right)$ with the base space $M$ representing spacetime, a 
gauge transformation is identical with the action of the
structure group $H$ along local fibres, being the spacetime base 
manifold not affected by such transformations.

Obviously, this scheme is not applicable to gauge theories of 
gravitation. In fact, they are theories of the Yang--Mills type 
based on the gauging of spacetime groups; and precisely these 
groups are symmetries which affect spacetime itself. Thus, one 
has to generalize the definition of gauge transformations in order 
to take account of such external symmetries as well. We follow 
Lord \cite{LO86,LO87,LG88}, who suppresses the restriction of no action 
on the base space. According to him, a gauge transformation is a 
general bundle automorphism, that is, a diffeomorphism that maps 
fibres to fibres. 

The natural framework to define such an action is the following,
based on the manifold character of Lie groups and on the simple 
properties of left and right multiplications of group elements. 
Let us choose a spacetime group $G$, and a subgroup $H\subset
G$. We will construct the gauge theory of $G$ on the principal
fibre bundle $G\left( G/H\,,H\,\right)$, where the group
manifold $G$ itself is the bundle manifold, and the subgroup $H$
is taken to be the structure group ---different $H$'s may be
chosen to play this role---; the quotient space $G/H$ will play 
the role of spacetime. Gauge transformations are defined on this 
bundle as follows. As mentioned above, the usual definition of 
(active) gauge transformations as vertical automorphisms along 
fibres, not affecting spacetime, must be modified to more 
general automorphisms affecting both, vertical fibres and the 
points of the quotient space $G/H$ the latter are attached to. 
Since the left and right multiplications of elements of $G$ commute, 
we have in particular $L_g\circ R_h=R_h\circ L_g$, with $g\in G$, 
$h\in H$. Thus $L_g$, acting on fibres defined as orbits of the right 
action $R_h$ (that is, as left cosets $gH$), constitutes an 
automorphism of the kind we are looking for, transforming in general 
fibres into fibres. To be explicit, we define the left action $L_g$ 
of $G$ on zero sections $\sigma :G/H\longrightarrow G$ as follows:
\be 
L_g\circ\sigma (\xi\,)=R_h\circ\sigma (\xi\,'\,)\,.\label{Lie}
\ee
As observed by Lord \cite{LO87}, this equation coincides with the
prescription for {\em nonlinear transformations} due to Coleman et
al. \cite{CWZ69}. Thus, the NLR-procedure rests on a particular definition (\ref{Lie}) of the action of a given group $G$. As a prerequisite, a subgroup $H\subset G$ is necessary in order to perform a partition of the group manifold of $G$ into equivalence classes, namely the coset spaces $gH$, playing the role of fibres attached to each point of $G/H$. The group action (\ref{Lie}) moves from a fibre to another. In accordance with what one expects for spacetime symmetries, a transformation is induced on the quotient space $G/H$, reflecting the mapping from a fibre to other. Indeed, taking into 
account that $\pi\circ R_h\circ\sigma =\pi\circ\sigma =id$, from 
(\ref{Lie}) then follows 
\be 
\xi '=\pi\circ L_g\circ\sigma (\xi\,)\,,
\ee 
being the fields $\xi $ {\em coset fields}, characterized as 
continuous labels of the elements of the quotient space $G/H$. In 
particular, for $G/H\approx R^4$, they provide the translational 
fields destinated to replace the `quartet' of scalar fields of 
Refs. \cite{GMMH98,gron97}. Actually, in the context of gauge 
theories of spacetime groups, the `Poincar\'e coordinates'\cite{GN92} 
or components \cite{Cartan} of ``Cartan's radius vector" $\xi ^\alpha$ 
are in fact translational coset fields. In the case of MAG we are 
interested in, they turn out to transform as affine covectors 
resembling coordinates, see (\ref{affcov}) and (\ref{affcovb}) below. 

\subsection{Nonlinear connections}

For practical calculational purposes, the fundamental Eq. (\ref{Lie}) defining the nonlinear group action may be rewritten in a more explicit 
form in terms of $g\in G$ and $h\in H$ as  
\be
g\,\sigma (\xi\,)=\,\sigma (\xi\,'\,)\,h\left(\xi\,, g\,\right)\label{gact}
\,,\ee
or shortly as $\sigma '=g\,\sigma\,h^{-1}$. Due to this particular 
transformation law of $\sigma$, from the linear connection ${\buildrel\approx\over\Gamma}$ of $G$ it becomes possible to define 
the nonlinear connection $\Gamma $ with suitable transformation 
properties as follows 
\be 
\Gamma :=\,\sigma ^{-1}\left( 
d\, +{\buildrel\approx\over\Gamma}\,\right)\sigma\,.
\label{nonlcom}
\ee 
Indeed, given the ordinary linear transformation law of the linear 
connection, ${\buildrel\approx\over\Gamma}$, namely 
\be
{\buildrel\approx\over\Gamma} '=\, g\,{\buildrel\approx\over\Gamma} 
\,g^{-1}+g\,d\,g^{-1}\,, \ee
and the transformation (\ref{gact}) of $\sigma$ in its shorted form 
$\sigma '=g\,\sigma\,h^{-1}$, it follows that the nonlinear connection (\ref{nonlcom}) transforms as
\be
\Gamma '=\,h \Gamma h^{-1} +h d\, h^{-1} \label{con}
\ee 
under local transformations. Observe that, according to (\ref{con}), 
only the components of $\Gamma$ defined on the Lie algebra of 
$H\subset G$ transform inhomogeneously as true connections; the remaining 
components of $\Gamma$ transform as tensors with respect to
$H$. 

The nonlinear connection allows to construct covariant derivatives 
(of nonlinear fields) as follows. Consider a field $\varphi$ 
transforming linearly under $G$ as $\varphi ' =g\varphi$, and let 
us schematically define a correlated nonlinear field as 
$\psi :=\sigma ^{-1}\,\varphi$. (For more details on the explicit form of this relation, regarding in particular multispinors, see \cite{LPTT96}.) It is trivial to test that $\psi$ transforms under the action of $G$ as $\psi ' =h\psi $, that is, as a representation field of the subgroup $H$. Accordingly, we define the covariant differential 
\be 
D\psi :=\,\left( d\, +\Gamma\,\right)\psi 
=\sigma ^{-1}\left( d \,+{\buildrel\approx\over\Gamma}\,\right)
\varphi\,, 
\ee 
behaving as an $H$--covariant object, namely 
\be 
\left( D\psi\,\right) ' =h D\psi\,, 
\ee 
under the left action of the whole group $G$.

\section{Gauge theoretical origin of the tetrads}
By  applying the previous results, taking $G$ to be the 
{\em affine group} $A(4\,,R):=R^{4} \semidirect GL(4\,,R) $, i.e. 
the semidirect product of translations and general linear transformations, we will show how the nonlinear approach leads to MAG. The commutation relations of the affine group are:
\ba
\left[P_{\alpha}\, , P_{\beta}\right] &=& 0,\nonumber \\ 
\left[L^{\alpha}{}_{\beta}\, ,  P_{\gamma}\right] &=&
\delta^{\alpha}_{\gamma}\, P_{\beta}\, , \nonumber\\
 \left[L^{\alpha}{}_{\beta},  L^{\gamma}{}_{\delta}\right] &=&
\delta^{\alpha}_{\delta}\, L^{\gamma}{}_{\beta} -
\delta^{\gamma}_{\beta}\, L^{\alpha}{}_{\delta}\, . 
\ea
Observe that the physical dimensions of the generators of the linear group  are
$[L^{\alpha}{}_{\beta}\, ]=\hbar$, whereas those of translations are
$[P_{\alpha}\, ]=\hbar/{\rm
length}$.

Let us construct in two steps the  fibre bundle descriptions of
the spacetime dynamics of $G=A(4\,,R)$, which we denote
$G\left( G/H_1\,,H_1\,\right)$ and $G\left( G/H_2\,,H_2\,\right)$, 
corresponding to two consecutive smaller subgroups as structure group, 
namely $H_1=GL(4\,,R)$ and $H_2=SO(1\,,3)\in H_1$ respectively. Other
choices of $H$ are possible, for instance $H_3=SO(3)$, see
Ref.\cite{LPTT97}, but this will not be considered here. The 
occurrence of a certain subgroup $H$ (or $H_1$, $H_2$ etc.) on which 
the action of the total group $G$ becomes projected, is a constitutive 
feature of NLRs; it should not be confused with {\em symmetry breaking}. 
Indeed, in the nonlinear approach the symmetry is not broken, so that 
alternative choices of the subgroup $H$ are mathematically equivalent. 
True symmetry breaking requires an additional mechanism involving the ground state of a dynamical theory of fundamental physics. Since we do not propose here such a symmetry breaking mechanism for the original group $G$, the gauge theories we are going to construct are to be seen as different realizations of the unique underlying gauge theory of the whole gauge group $G$ (in our case, the affine group). The subgroups $H\subset G$ (in particular, $H=$Lorentz) do not play a singularized dynamical role. In fact, different choices of $H_1\,,H_2\subset G$ are interchangeable, in the sense that one can go from a realization $G\left( G/H_1\,,H_1\,\right)$ to another $G\left( G/H_2\,,H_2\,\right)$, being both, simply, different rearrangements of the degrees of freedom of the same gauge theory, namely MAG. The possibility of dynamically singling out one and only one of the subgroups $H\subset G$, say the Lorentz group, by means of a symmetry breaking mechanism, remains to be studied elsewhere. 

{}First we consider the gauge theory of the affine group with
 the general linear group $H_1=GL(4\,,R)$ as structure group. We
will show that the coframe appears in a natural way as a
nonlinear translative connection.

Now in the formula (\ref{gact}) for the nonlinear group action, 
we substitute the following quantities: The  group 
elements $g$ of the whole affine group $A(4\,,R)$ are parametrized as  
\be
g=\,  e^{i\,\epsilon ^{\alpha} P_\alpha}
e^{i\,\omega _\alpha {}^\beta L  ^\alpha {}_\beta}
\approx I+i\,\epsilon ^{\alpha} P_\alpha 
+i\,\omega _\alpha {}^\beta L  ^\alpha {}_\beta\,,\label{param}
\ee
where we also indicate the infinitesimal expansion.
They act on the zero sections 
\be
\tilde{\sigma}(\xi\,):=e^{-i\,\xi ^\alpha P_\alpha}\,,\label{zerosec}
\ee
where $\xi ^\alpha$ are (finite) coset parameters. We introduce
the tilde in order to distinguish (\ref{zerosec}) from the $\sigma$
introduced in (\ref{fact}) below. The  elements $h$ of the
structure group $GL(4\,,R)$ are taken to be 
\be
h:=e^{i\,v _\alpha {}^\beta L  ^\alpha {}_\beta}
\approx I+i\,v _\alpha {}^\beta L  ^\alpha {}_\beta
\,. \label{infh}
\ee
Using the Campbell--Hausdorff formula in (\ref{gact}) 
with (\ref{param}--\ref{infh}), the 
variation of the coset parameters $\xi ^{\alpha}$ of (\ref{zerosec}) and
the value of $v _\alpha {}^\beta$, see (\ref{infh}), are calculable resulting in 
\be
\delta\xi ^{\alpha }=-\omega _\beta {}^\alpha
\,\xi ^\beta -\epsilon ^\alpha\quad\,,\qquad v _\alpha {}^\beta 
=\,\omega _\alpha {}^\beta \,.\label{affcov}
\ee
Thus we see from (\ref{affcov}) that the coset parameters $\xi ^\alpha$ 
transform as affine covectors, as postulated \cite{Cartan} for {\em Cartan's
generalized radius vector}. The nonlinear connection (\ref{nonlcom}) will be
constructed in terms of the linear one, namely 
\be 
{\buildrel\approx\over\Gamma} :=-i\,{\buildrel ({\rm T})\over{\Gamma ^\alpha}} P_\alpha 
         -i\,{\buildrel ({\rm GL})\over{\Gamma _\alpha {}^\beta }}
         L  ^\alpha {}_\beta \,,\label{lincon}\ee
which includes the linear translational potential 
${\buildrel ({\rm T})\over{\Gamma ^\alpha}}$ and the $GL(4\,,R)$
connection ${\buildrel ({\rm GL})\over{\Gamma _\alpha {}^\beta }}$,
whose infinitesimal transformations read 
\be 
\delta {\buildrel ({\rm GL})\over{\Gamma _\alpha {}^\beta }}=\,
{\buildrel ({\rm GL})\over{D}}\omega _\alpha {}^\beta\,,\qquad
\delta {\buildrel ({\rm T})\over{\Gamma ^{\alpha}}} =\,
{\buildrel ({\rm GL})\over{D}}\epsilon ^\alpha 
-\omega _\beta {}^\alpha 
{\buildrel ({\rm T})\over{\Gamma ^{\beta}}}\,.\label{lintr}
\ee
Here ${\buildrel ({\rm GL})\over{D}}$ denotes the covariant differential 
constructed from  the $GL(4\,,R)$ connection. Making use of the
definition (\ref{nonlcom}), we get
\be
\tilde{\Gamma }:=\,\tilde{\sigma}^{-1}\left(d\,+{\buildrel\approx\over\Gamma}
\,\right)\tilde{\sigma} =-i\,\tilde{\vartheta }
^\alpha P_\alpha -i\,\tilde{\Gamma }_\alpha {}^\beta L  
^\alpha {}_\beta\,.\ee
with
\be
\tilde{\Gamma }_\alpha {}^\beta ={\buildrel ({\rm GL})
\over{\Gamma _\alpha {}^\beta }}\,,\qquad 
\tilde{\vartheta }^\alpha :={\buildrel ({\rm T})\over{\Gamma
^\alpha}} +{\buildrel ({\rm GL})\over D}\xi ^\alpha
\,. \label{deftetr}
\ee
As in the case of (\ref{zerosec}), we denote these objects with a tilde 
for later convenience. Making use of (\ref{con}), it is straightforward 
to prove that, whereas $\tilde{\Gamma}_\alpha {}^\beta$ transforms as a $GL(4\,,R)$ connection, the coframe $\tilde{\vartheta }^\alpha$ defined as in (\ref{deftetr}) transforms as a $GL(4\,,R)$ covector. Explicitly 
\be 
\delta \tilde{\Gamma }_\alpha {}^\beta =\,
\tilde{D}\omega _\alpha {}^\beta\,,\qquad
\delta \tilde{\vartheta }^{\alpha} =-\omega _\beta {}^\alpha \tilde{\vartheta }^{\beta}\,,
\ee
compare with (\ref{lintr}). The nonlinear treatment of the affine group 
thus clarifies how the coframe can be constructed from gauge fields of 
the Yang--Mills type, in particular those of (\ref{lincon}). The coset parameters $\xi ^\alpha$ play the role of Cartan's generalized radius 
vector of Ref.\cite{MMNH93}, being not introduced {\em ad hoc}, since 
they are constitutive elements of the theory. They mainly contribute to 
the construction of the translational invariant $\tilde{\vartheta}^\alpha 
=\,{\buildrel {({\rm T})}\over{\Gamma ^\alpha}}+\tilde{D}\xi ^\alpha$; 
the variation of $\xi$ under translations, see (\ref{affcov}), is 
compensated by the variation of the translative connection, see (\ref{lintr}), cf. \cite{MMNH93}. Since $\xi=\xi^{\alpha}P_{\alpha}$ 
aquires its values in the coset space $A(n,R)/GL(n,R)\approx R^n$, it can be regarded as an affine vector field (or ``generalized 
Higgs field" according to Trautman \cite{Tr79}) 
which ``hides" \cite{[28]} the action of the local translational symmetry ${\cal T}(n,R)$. Accordingly, conditions like ${\buildrel {({\rm T})}\over{\Gamma ^\alpha}} =0$ or 
$D\xi ^\alpha =0$ break the translational symmetry. Only in the
absence of gravitational interaction we recover the specially
relativistic relation $\vartheta ^\alpha =\,d\xi ^\alpha $ for the 
coframes (i.e., for the translational nonlinear connections), 
employed in Ref. \cite{GMMH98} in order to derive a ``metric--free" 
volume four-form. It is interesting to notice that, in this limit, 
the fields $\xi ^\alpha$ play the role 
of ordinary coordinates, see also (\ref{affcov}). In other words, 
the spacetime manifold of special 
relativity is a {\em residual} structure of the nonlinear approach 
when gravitational forces are switched off.

\section{The metric in MAG}
 In order to complete the MAG scheme, the next  considerations are devoted
to show how the metric tensor can be introduced, related to a particular choice of the subgroup $H\subset G$. 

Let us consider the second choice of structure subgroup in our 
bundle approach mentioned above, namely $G\left( G/H_2\,,H_2\,
\right)$ with $G=A(4\,,R)$, as before, and $H_2=SO(1\,,3)$. We 
split up the generators $L^\alpha {}_\beta $ of the general linear transformations as $L  ^\alpha {}_\beta =\,{\buildrel\circ\over L}
{} ^\alpha {}_\beta + S ^\alpha {}_\beta$, being ${\buildrel\circ
\over L}{} ^\alpha {}_\beta$ the Lorentz generators and $S ^\alpha 
{}_\beta $ those of the symmetric linear transformations. Now we 
apply the general formula (\ref{gact}) with the particular 
factorization
\be 
g=\,  e^{i\,\epsilon ^{\alpha} P_\alpha}
e^{i\,\alpha ^{\mu\nu }S_{\mu\nu}}
e^{i\,\beta ^{\mu\nu }{\buildrel\circ\over L}_{\mu\nu}}\quad\,,\quad
\sigma :=e^{-i\,\xi ^\alpha P_\alpha}e^{i\,h^{\mu\nu }S_{\mu\nu}}
\quad\,,\quad h:=e^{i\,u ^{\mu\nu }{\buildrel\circ\over L}_{\mu\nu}}\quad\,.
\label{fact}\ee

Being $\epsilon ^{\alpha }$, $\alpha ^{\mu\nu }$ and $\beta
^{\mu\nu }$ infinitesimal parameters of the affine group, the
transformed coset parameters of $\sigma$ reduce to $\xi 
^{'\alpha} =\,\xi ^{\alpha} +\delta\xi ^{\alpha}$ and 
$h^{'\mu\nu }=\,h^{\mu\nu } +\delta h^{\mu\nu }$; the Lorentz
parameters $u ^{\mu\nu }$ (being Lorentz the structure group
$H_2$) are also infinitesimal. Let us define 
\be 
r_\alpha{}^\beta :=\,\left( e^h\right) _\alpha{}^\beta 
:=\delta _\alpha{}^\beta +h_\alpha{}^\beta +{1\over{2!}}
h_\alpha{}^\gamma h_\gamma{}^\beta + \cdots\label{rtrans}
\ee 
from the coset parameters $h^{\alpha\beta }$ associated to the 
generators of the symmetric part of $GL(4\,,R)$, see (\ref{fact}). 
(In the following, $r^{\alpha\beta }$ rather than the coset 
parameters $h^{\alpha\beta }$ themselves, will play the fundamental role\cite{BO74,LPTT95}.) We find the variations 
\be 
\delta\xi ^{\alpha }=-\left(\alpha _\beta {}^\alpha 
+\beta _\beta {}^\alpha \right) 
\,\xi ^\beta -\epsilon ^\alpha\,,\qquad
\delta r^{\alpha\beta }=\,\left(\alpha ^\alpha {}_\gamma 
+\beta ^\alpha {}_\gamma\right)\, r^{\gamma\beta }+ 
u^\beta {}_\gamma\, r^{\gamma\alpha }\,.\label{affcovb}
\ee
where $\alpha _\beta {}^\alpha +\beta _\beta {}^\alpha 
=\omega _\beta {}^\alpha$, compare with (\ref{affcov}). Since 
$r^{\alpha\beta }$ is symmetric, the antisymmetric part of the
second equation in (\ref{affcovb}) vanishes. {}From this condition 
we find the explicit form of the nonlinear Lorentz parameter
\be 
u^{\alpha\beta }=\,\beta ^{\alpha\beta } 
-\alpha ^{\mu\nu }\tanh\left\{ {1\over 2}
\log\left[ r^\alpha {}_\mu\,\left( r^{-1}\right) ^\beta {}_\nu
\right]\right\}\,,
\label{nonLo}
\ee
which obviously differs from the linear Lorentz parameter 
$\beta ^{\alpha\beta }$. It is precisely the nonlinear 
$u^{\alpha\beta }$, and not the linear $\beta ^{\alpha\beta }$, 
 which is relevant for nonlinear transformations, 
as becomes evident in (\ref{Locon}), (\ref{Locov}) and (\ref{trnm}) 
below.

In order to define the nonlinear connection, let us first rewrite 
the linear affine connection (\ref{lincon}) as 
\be 
{\buildrel\approx\over\Gamma} :=-i\,{\buildrel ({\rm T})\over{\Gamma ^\alpha}} P_\alpha 
         -i\,{\buildrel ({\rm GL})\over{\Gamma _\alpha {}^\beta }}
  \left( S^\alpha {}_\beta +
  {\buildrel\circ\over L}{}^\alpha {}_\beta \right)\,.\ee
The symmetric part of ${\buildrel ({\rm GL})\over{\Gamma _\alpha {}^\beta }}$ will correspond to nonmetricity. Making use of the definition (\ref{nonlcom}), we get
\be
\Gamma :=\,\sigma ^{-1}\left(d\,+{\buildrel\approx\over\Gamma}\,\right)\sigma =-i\,\vartheta
^\alpha P_\alpha -i\,\Gamma _\alpha {}^\beta \left( S^\alpha {}_\beta 
+{\buildrel\circ\over L}{}^\alpha {}_\beta\right)\,,\ee
with the nonlinear $GL(4\,,R)$ connection $\Gamma _\alpha
{}^\beta $ and the nonlinear translational connection $\vartheta 
^\alpha $ respectively defined as
\be
\Gamma _\alpha {}^\beta :=\,\left( r^{-1}\right)_\alpha {}^\gamma 
{\buildrel ({\rm GL})\over{\Gamma _\gamma {}^\lambda }}
\, r_\lambda {}^\beta -\left( r^{-1}\right) 
_\alpha {}^\gamma \, d\,r_\gamma {}^\beta\,,\qquad 
\vartheta ^\alpha :=\,r_\beta {}^\alpha
\left( {\buildrel ({\rm T})\over{\Gamma ^\beta}}
+{\buildrel ({\rm GL})\over{D}}\xi ^\beta \right)\,.
\label{noncon}
\ee
We identify the components of $\Gamma _\alpha {}^\beta $ and 
$\vartheta ^\alpha $ of the (nonlinear) Yang--Mills connections of
the affine group with the {\it{geometrical}} linear connection and 
with the coframe respectively. Thus Eqs. (\ref{noncon}) establish the 
correspondence between the {\it{geometrical}} objects in the
l.h.s. and the dynamical objects in the r.h.s. We find that the whole connection, consisting of the sum of the antisymmetric (Lorentz) part and the symmetric (nonmetricity) part, behaves as a Lorentz connection
\be
\delta \Gamma _\alpha {}^\beta =\,D\,u_\alpha {}^\beta
\,\label{Locon}
\ee
according to (\ref{con}), with the nonlinear Lorentz parameter (\ref{nonLo}). The transformation of the symmetric part of 
$\Gamma _\alpha {}^\beta $ lacks an inhomogeneous contribution, see (\ref{trnm}) below; it becomes a tensor as a result of the nonlinear realization. Let us mention that, in addition, the covariant 
differential $D$ in (\ref{Locon}) is constructed in terms of 
the Lorentz connection itself. On the other hand, the coframe 
transforms as a Lorentz covector 
\be 
\delta\vartheta ^{\alpha }
=-u_\beta {}^\alpha \vartheta ^\beta\,,\label{Locov}
\ee 
which constitutes a main result of the nonlinear approach. 

Notice that, in view of the splitting of the general linear 
generators into a Lorentz plus a symmetric part as 
$L^\alpha {}_\beta =\,{\buildrel\circ\over L}{} ^\alpha 
{}_\beta + S ^\alpha {}_\beta$, the connection is actually 
composed of two parts, defined on different elements of the Lie 
algebra. In fact, only the antisymmetric part, defined on the 
Lorentz generators, behaves as a true connection of the Lorentz 
group playing the role of the structure group $H_2$. As already mentioned, the symmetric part $\Gamma _{(\alpha\beta )}=:{1\over 2}Q_{\alpha\beta}$, 
i.e. the nonmetricity, is tensorial. Actually 
\be 
\delta Q_{\alpha\beta}=2\,u_{(\alpha}{}^\gamma
Q_{\beta )\gamma}\,.\label{trnm}
\ee 
On the other hand, being the structure group $H_2$ the Lorentz 
group, the Minkowski metric $o_{\alpha\beta }$ is automatically present in the theory as a natural invariant: $\delta o_{\alpha\beta }=\,0$. Thus, a metric occurs in the affine theory due to the particular choice of a (pseudo-)orthogonal group as the structure group, so that the corresponding Cartan-Killing metric becomes apparent. However, no degrees of freedom are related to the Minkowski metric. This seemingly makes a difference between 
the dynamical content of our theory and that of ordinary MAG, since 
in the latter the metric tensor involves ten degrees of freedom. 
Nevertheless, we will see immediatly how these degrees of freedom, 
being of Goldstone nature, can be taken from the nonlinear 
connections where they are hidden. Actually, the Goldstone fields 
which will manifest themselves as the degrees of freedom of the 
MAG--metric are those of the matrix $r^{\alpha\beta }$ defined in (\ref{rtrans}). They can be factorized into the nonlinear 
connections and the coframes, as shown in (\ref{noncon}), in the 
presence of the Minkowskian metric we are discussing, or alternatively 
they can be explicitly displayed in the metric tensor, as we will 
show in (\ref{metr}) below. In this case, the metric becomes identical 
with the ordinary MAG metric. 

In order to show how the transition between these alternative 
formulations takes place, we establish the correspondence 
between the objects of both choices $H_1$ and $H_2$ studied 
above. Formally, we find that this correspondence is isomorphic 
to a finite gauge transformation, with the matrix $r^{\alpha\beta }$ 
of (\ref{rtrans}) standing for the symmetric affine transformations. 
But $r^{\alpha\beta }$ is not a transformation matrix; it is 
constructed in terms of coset fields. The relation between (\ref{deftetr})and (\ref{noncon}) reads
\be
\tilde{\Gamma }_\alpha {}^\beta :=
\,{\buildrel ({\rm GL})\over{\Gamma _\alpha {}^\beta }}
=\,r_\alpha {}^\gamma
\Gamma _\gamma {}^\lambda\left( r^{-1}\right)_\lambda {}^\beta
-r_\alpha {}^\gamma d\,\left( r^{-1}\right)_\gamma {}^\beta
\,,\label{corra}\ee
and 
\be
\tilde{\vartheta }^\alpha :=\,
{\buildrel ({\rm T})\over{\Gamma ^\alpha}}
+{\buildrel ({\rm GL})\over{D}}\xi ^\alpha 
=\, \left( r^{-1}\right) _\beta {}^\alpha \vartheta ^\beta
\,.\label{corrb}\ee
The standard metric--affine objects of ordinary MAG, such as 
connections and coframes (up to the metric), are identical to 
those with tilde in the l.h.s. of (\ref{corra}) and (\ref{corrb}), 
studied in Sec.  III, corresponding to a nonlinear realization 
of the affine group with $H_1=GL(4\,,R)$ as the structure group. 
In the approach studied in Sec.  III, the metric tensor was 
absent. However, in analogy to (\ref{corra}) and (\ref{corrb}), 
it can be introduced as an object with tilde related to the 
Minkowski metric $o_{\alpha\beta }$ which appears in 
the case of $H_2=SO(1\,,3)$ studied in Sec.  IV. Actually, 
we define $\tilde{g}_{\alpha\beta }$ from $o_{\alpha\beta }$ as 
\be
\tilde{g}_{\alpha\beta }:=\,r_\alpha {}^\mu r_\beta {}^\nu 
o_{\mu\nu }\,.\label{metr}
\ee
The resulting MAG-metric tensor plays the role of a {\em Goldstone 
field}, cf. \cite{CWZ69}, that  drops out after applying the 
inverse of the "gauge transformation" (\ref{metr}). By also inverting 
(\ref{corra}) and (\ref{corrb}), one reaches the nonlinear 
realization studied in Sec.  IV, with the Lorentz group as the structure subgroup. This completes the correspondence between the nonlinear objects and those of the framework of ordinary metric--affine theory. The virtue of eqs.(\ref{corra}--\ref{metr}) consists in that they show explicitly the mechanism by means of which the rearrangement of gauge-theoretical degrees of freedom (namely those related to the symmetric general linear transformations) takes place by "choosing tetrads to be orthonormal" in the context of MAG. As a consequence, observe that invariants like the line element may be alternatively expressed in terms of the Lorentz-nonlinear or metric--affine objects respectively, namely as
\be
ds^2=\,o_{\alpha\beta }\vartheta ^\alpha\otimes\vartheta ^\beta 
=\,\tilde{g}_{\alpha\beta }\tilde{\vartheta }^\alpha
\otimes\tilde{\vartheta }^\beta \,,
\label{metric}
\ee
where the transition from $o_{\alpha\beta }$ to 
$\tilde{g}_{\alpha\beta }$ or {\em vice versa} takes place by 
means of the suitable factorization of the coset parameters 
associated to the symmetric affine transformations. Thus, given the standard MAG theory, if one fixes the metric to be globally Minkowskian, the degrees of freedom of the theory automatically rearrange themselves into the nonlinear theory developed in Sec. IV, with the Lorentz group as the
structure group. However, the theory remains the same. 

Due to the transformation law (\ref{affcovb}) of $r^{\alpha\beta}$, 
which involves both, general linear and Lorentz parameters, the
indices of objects with tilde behave as general linear
indices; those of objects without tilde are Lorentz indices. In
the second case, the ten degrees of freedom corresponding to
$r^{\alpha\beta}$ are rearranged into the coframe and
connections, so that none of them remains in the metric tensor,
which becomes Minkowskian. An action which is invariant under 
affine transformations can be alternatively expressed in terms
of $GL(4\,,R)$ or $SO(1\,,3)$ tensors, respectively. This 
corresponds to the choice of variables with or without tilde, as
discussed above. The field equations derived, by means of a variational principle, from such an affine invariant action, are the same as those derived in \cite{hehl95}. As already known by the authors of the latter reference, the field equation obtained by varying with respect to the MAG metric tensor is redundant, in accordance with the fact that the Goldstone-like degrees of freedom of such quantity can be rearranged into the remaining fields of the theory. Being the metric tensor eliminable as a fundamental gravitational potential, the interpretation of nonmetricity as the corresponding field strength can be put aside. In the nonlinear approach, it manifests itself simply as the (tensorial) symmetric part of the linear connection, so that action terms quadratic in nonmetricity should be seen as mass terms rather than as kinetic terms. This alternative conceptual point of view leaves the theory formally unchanged.

\section{Outlook: Dynamical origin of the signature?}

In the present work, we were not concerned with the coupling of the gravitational gauge theory to matter. We refer the interested reader to Ref.\cite{LPTT96}, where the relation between multispinors and ordinary spinors was studied. Such relation extends the correspondence between different realizations of the affine group, displayed in eqs.(\ref{corra}--\ref{metr}), to matter fields. Although the details of the coupling to dilatation and shear currents were not worked out, in principle they could be obtained from the correspondence established there.  

Concerning the {\em signature} of the metric parametrized via 
$o_{\alpha\beta}:= {\rm diag}(e^{i\theta}, 1, 1, 1)$, cf. \cite{ORT94},
the nonlinear approach is particularly adapted for dealing with {\em spontaneous symmetry breaking}. In fact, the Higgs mechanism can be understood as the way to select a particular structure group $H$ by 
fixing the Goldstone fields in terms of suitable fields of the theory, 
see Ref. \cite{TT98}. Thus, symmetry breaking could give a fundamental physical meaning to a particular structure subgroup $H$, fixing it 
dynamically. 

Previously to the symmetry breaking, the choices of 
different structure groups $H$ are physically equivalent in the sense 
that they simply provide alternative ways to rearrange the degrees of freedom of the total gauge group $G$. In particular, in the gauge theory 
of the affine group, in the absence of symmetry breaking one can 
freely choose the structure subgroup $H$ to be the Lorentz group 
or $SO(4)$ etc., so that the corresponding metric signatures 
become the Minkowskian or the Euclidean one, respectively. They constitute alternative realizations of the same theory, since it is the symmetry 
under the total group $G$ the only relevant one. 

In quantum field theory, Minkowskian or Euclidean signature are, however, quite different. Usually in the path integral approach, Euclidean 
signature is chosen in order to have a well defined measure. 
Moreover, tunneling between different topologies of instanton 
configurations may occur. 
(After applying a Wick rotation to $e^{i\pi}=-1$, the physical 
measurable quantities are regained.) In the path integral approach 
to quantum gravity, a summation over all {\em inequivalent} coframes and connections, 
and even topology \cite{Mi77} is understood. This summation will also 
involve {\em degenerate} ($det\, e_{j}{}^{\beta}=0$) or even 
vanishing coframes, cf. \cite{Ja99}. Macroscopically, this would 
imply the breakdown of any length measurement performed by means 
of the metric (\ref{metric}). Microscopically, then also signature 
changes of the metric are to be admitted, cf. Refs. \cite{E92,DT93}. 
These conceptual difficulties \cite{Is91} are not encountered in 
the quantization of internal 
Yang--Mills theories on a {\em fixed} spacetime background.

Degenerate coframes, however, tend to jeopardize the coupling
of gravity to matter fields, as exemplified by Dirac or Rarita--Schwinger
fields, cf. \cite{MM99}. The basic reason is that the local frame $e_{\alpha}$, even if it still exists, is not invertible any more; 
i.e. the relation 
$e_{\alpha}\rfloor\vartheta^{\beta}=\delta_{\alpha}^{\beta}\>$, which
is needed in the formulation of matter Lagrangians, would then be lost. 

These arguments seems to require to introduce a 
symmetry fixing mechanism which dynamically differentiates 
a particular structure group $H$, and thus the signature. 
In other words, it remains to be seen if also the signature of 
the physical spacetime has a dynamical origin in such a framework, 
as is suggested by Sakharov \cite{[43]} and  Greensite \cite{[42]}, 
or arises naturally in string or M-theory \cite{W98,Du99}.

 \section*{Acknowledgments} 
We would like to thank Friedrich W. Hehl, Alfredo Mac\'ias, Yuri 
Obukhov and Alfredo Tiemblo for useful hints and comments.
This work was partially supported by  CONACyT, grant No. 28339--E, 
and the joint German--Mexican project DLR--Conacyt
 E130--1148 and MXI 010/98 OTH. One of us (E.W.M.)  
thanks Noelia M\'endez C\'ordova for encouragement.

\end{document}